# Combining Reinforcement Learning with Graph Convolutional Neural Networks for Efficient Design of TiAl/TiAlN Atomic-Scale Interfaces

Xinyu Jiang, Haofan Sun, Qiong Nian[†] and Houlong Zhuang*

School for Engineering of Matter, Transport and Energy, Arizona State University, Tempe, AZ 85287, USA

[†]Qiong.Nian@asu.edu

* zhuanghl@asu.edu

**Abstract**

Ti/TiN coatings are utilized in a wide variety of engineering applications due to their superior properties such as high hardness and toughness. Doping Al into Ti/TiN can also enhance properties and lead to even higher performance. Therefore, studying the atomic-level behavior of the TiAl/TiAlN interface is important. However, due to the large number of possible combinations for the 50 mol% Al-doped Ti/TiN system, it is time-consuming to use the DFT-based Monte Carlo method to find the optimal TiAl/TiAlN system with a high work of adhesion. In this study, we use a graph convolutional neural network as an interatomic potential, combined with reinforcement learning, to improve the efficiency of finding optimal structures with a high work of adhesion. By inspecting the features of structures in neural networks, we found that the optimal structures follow a certain pattern of doping Al near the interface. The electronic structure and bonding analysis indicate that the optimal TiAl/TiAlN structures have higher bonding strength. We expect our approach to significantly accelerate the design of advanced ceramic coatings, which can lead to more durable and efficient materials for engineering applications.

# 1. Introduction

Metal/ceramic coatings play a pivotal role across various fields, including wear-resistant materials, electronic devices, gas-turbine engines, and medical implants, by improving the structural and electrical properties of substrate materials [1-5]. The metal components that possess high toughness are combined with ceramic components [6], which exhibit high hardness, abrasion resistance, chemical stability, and wear resistance [7, 8], to reduce the trade-off between ductility and hardness.

Transition metal nitrides or carbide-based metal/ceramic coatings are commonly studied due to their technical feasibility [9]. Among these coatings, Ti/TiN coatings are extensively investigated in different engineering fields such as structural applications, biomedical implants, fuel cells, aircraft, and gas turbine engines owing to high hardness and toughness, particle erosion resistance, corrosion resistance, and tribocorrosion resistance [10-19]. Many studies have highlighted that by employing nano-scale multilayers, Ti/TiN multilayer coatings have higher toughness and high adhesive strength [20], higher erosion and corrosion resistance [14, 15, 17-19, 21]. Moreover, it is well known that Al can be doped into conventional binary transition metal nitride, such as TiN, to improve the properties and impact the adhesion of coating [22-26]. For example, compared to Ti/TiN coatings, TiAl/TiAlN coatings have higher hardness and elastic modulus [27], TiAl/TiAlN multilayer coatings also have good corrosion resistance and high adherence strength [28], and nanolayered multilayered TiAl/TiAlN coatings have high solid particle erosion and corrosion resistance [11]. Consequently, understanding the impact of dopants like Al on the mechanical properties of nanoscale

Ti/TiN coatings is imperative.

The quality of the interface significantly impacts mechanical properties in nanoscale multilayer coatings [8, 29]. A comprehensive understanding of the metal/ceramic interfaces is therefore necessary to design coatings with the desired strength [30]. However, shedding light on the complicated details of metal/TiN interfaces remains a formidable challenge [24, 31, 32]. To address this challenge, first-principles density functional theory (DFT) and molecular dynamics (MD) simulations emerge as powerful tools for unraveling the atomic-level behavior of metal/ceramic interfaces [33-36]. For example, the impact of Al doping on the shear strength of the Ti/TiN interface has recently been studied by DFT and MD simulations [31, 37], which show that the introduction of Al significantly increases the shear strength of the Ti/TiN interface. In these studies, a Monte Carlo (MC) approach is used to find low-energy doped configurations. However, this approach is time-consuming because it requires DFT calculations to relax each generated potential configuration, and both works studied interfaces with low Al compositions of up to 25 mol%. Since the introduction of Al to TiN has resulted in higher hardness up to 60 mol% [27, 38, 39], and the number of potential configurations increases significantly with the increase of doping atoms, it is necessary to find a new way to efficiently explore the shear strength of TiAl/TiAlN interface with higher Al compositions.

Machine learning (ML) method is an active area in material science because it has shown the potential to predict materials properties from their atomic structures or create atomic structures from desired properties [40, 41]. Among ML methods, graph-based neural networks have been successfully applied to materials property prediction and interatomic potential development due to their highly accurate prediction of DFT-calculated properties [42-45]. In addition, as optimization methods, reinforcement learning (RL) and genetic algorithm (GA) are widely used in designing different materials. For example, RL has been used in two-dimensional (2D) stretchable material design, interatomic potential development, chemical reaction optimization, and molecule discovery [46-49], and GA has been applied to interatomic-potential parameters optimization, feature selection, training configuration generation, and experimental parameters optimization due to its efficiency and stability [50-53]. Therefore, to efficiently obtain stable adhesive TiAl/TiAlN configurations and account for the periodic boundaries in crystals, we use crystal graph convolutional neural networks (CGCNN) as our interatomic potential to predict the work of adhesion ($W_{ad}$). We also combine CGCNN with different methods, such as Monte Carlo (MC), deep Q-network (DQN) [54], and genetic algorithms (GA), to identify candidates with high $W_{ad}$ and compare their performance. In this paper, we mainly focus on 50 mol% Al dopants. By analyzing the features of various layers within the CGCNN and DQN, we find that structures with high $W_{ad}$ exhibit a specific pattern of doping aluminum (Al) atoms near the interface. Further examination of the electronic structure and bonding in TiAl/TiAlN structures with both high and low $W_{ad}$, as well as in Ti/TiN structures, reveals that differences in atomic size and electronegativity significantly influence the bonding strength at the interface.

**Methodology**

**1.1 DFT calculations**

The Vienna *Ab initio* Simulation Package (VASP; version 5.4.4) [55-57] is used for the DFT calculations with the generalized gradient approximation (GGA) parameterized by Perdew, Burke, and Ernzerhof (PBE) [58]. The standard projector augmented wave (PAW) pseudopotentials are used for core electrons as implemented in VASP [59, 60], and plane wave basis sets with cutoff energies of 400 eV and 550 eV are utilized for non-ionic update calculations and ionic relaxation, respectively [61, 62]. A 12×9×1 mesh centered at Gamma is used for the metal/ceramic systems with periodic boundary conditions [63].

**1.2 TiAl/TiAlN interface structure and work of adhesion**

Fig. 1 shows the geometry of the Al-doped Ti/TiN structure, which consists of a total of 56 atoms (20 for TiAl and 36 for TiAlN). This structure is based on the Ti/TiN configuration, consisting of metal (Ti) and ceramic (TiN) layers, with the preferred crystallographic orientation based on the edge-to-edge matching model [64, 65]. Here, the face-centered cubic/hexagonal closed-packed (FCC/HCP) interface orientation relationship is chosen to create a single Ti (0 0 0 1)/TiN (1 1 1) interface, aligned normally to the z direction, which has the lowest interatomic spacing misfit, so, as shown in Fig. 1, the [$\bar{1}\bar{1}20$] and [$1\bar{1}0$] directions are parallel to the x direction, and the [$11\bar{2}$] and [$1\bar{1}00$] directions are set to parallel to the y direction. The in-plane lattice constants for these two different phases extracted from the Materials Project [66] are 2.94 Å and 5.08 Å for Ti and 3.00 Å and 5.19 Å for TiN. The mismatches in the interface are 2.0% and 2.2% in the x and y directions, respectively, indicating that the interface can be treated as a coherent interface where the Ti layer is stretched in both the x and y directions to match TiN before binding Ti and TiN layers together. To achieve strong adhesion to the Ti layer, N atoms in the TiN layer are connected to the Ti layer [8]. In this work, we focus on a 50 mol% composition of Al dopants in both layers, implying the exchange of 10 out of 20 Ti atoms in the Ti layer and 9 out of 18 Ti atoms in the TiN layer with Al atoms (see Fig. 1). A vacuum of 16-Å thickness is applied in the z direction to remove the interactions of the surface atoms between interface slab models.

When the interface is built, the $W_{ad}$ defined as the reversible work needed to separate the interface into two free surfaces [32, 33], can be a qualitative indicator of the binding strength. A higher $W_{ad}$ means stronger and energetically favorable binding strength [67, 68]. The $W_{ad}$ of the interface can be given by:

$$W_{ad} = (E_{TiAl} + E_{TiAlN} - E_{TiAl/TiAlN})/A \qquad (1)$$

where $E_{TiAl}$ and $E_{TiAlN}$ are the relaxed energy of TiAl and TiAlN isolated systems in contact with vacuum, respectively. after full relaxation, respectively; $E_{TiAl/TiAlN}$ is the total energy of the relaxed interface system. $A$ is the cross-sectional area of the interface slab model. When the area of the interface for all interface models is treated as a constant, an approximate $W_{ad}$, $W^{app}$ (Eq. 2), which removes the divisor interface area is applied in the CGCNN model to represent the bond strength at the interface:

$$W^{app} = E_{TiAl} + E_{TiAlN} - E_{TiAl/TiAlN} \qquad (2)$$

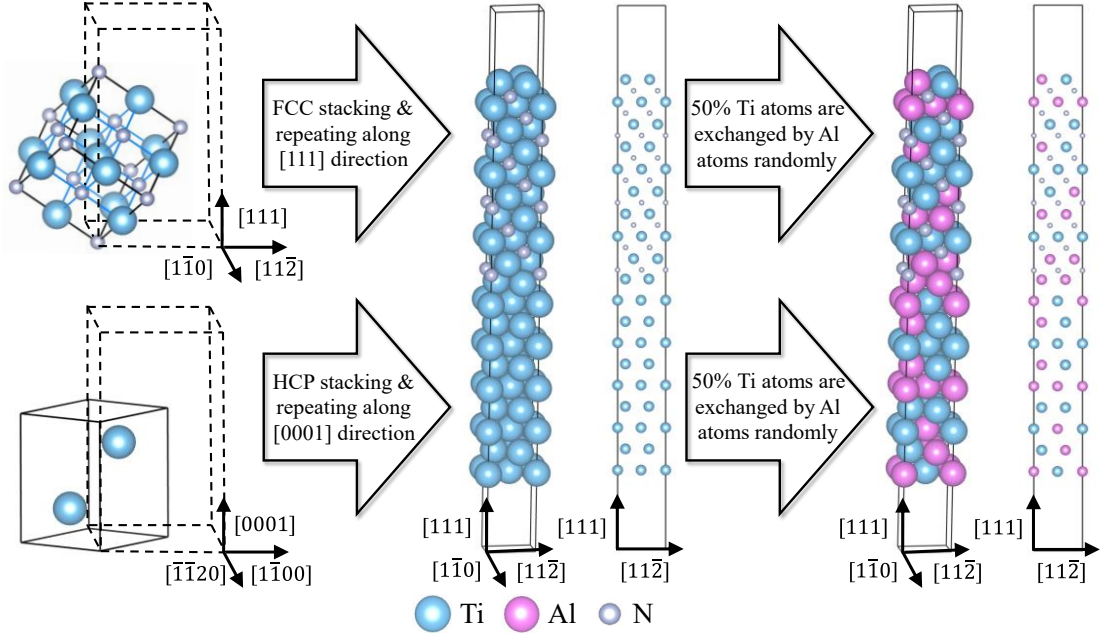

**Fig. 1.** Al-doped Ti/TiN interface structure with 50 mol% of Al generated from FCC/HCP Ti/TiN interface. Here, the directions of both Ti/TiN and TiAl/TiAlN interfaces are referenced by the TiN coordinate. Each system is shown by a perspective space-filling model and a 2D ball model.

**1.3 Crystal graph convolutional neural networks (CGCNN) for work of adhesion prediction**

For the 56-atoms Ti/TiN structure, alternating half of Ti atoms with Al atoms in both metal and ceramic parts results in C(18, 9) × C(20, 10) = 8,982,836,720 possible TiAl/TiAlN combinations. Since the total elapsed time of DFT calculation of $W_{ad}$ for each TiAl/TIAlN structure is more than one day with 64 computer cores, it is impossible to search such vast structure space to find potential structure by DFT relaxation of structures from random search. MC approach can expedite the exploration process, but it still needs time-consuming DFT calculation to relax each potential structure [31, 37]. To further enhance the efficiency, the CGCNN model instead of DFT calculation is utilized here. By randomly exchanging half Ti atoms by Al atoms in each part of the Ti/TiN system to get TiAl/TiAlN systems, the $W^{app}$ of generated TiAl/TiAlN structures is computed by using DFT no ionic update calculations and feeding these $W^{app}$ and corresponding structures into CGCNN to train the model.

CGCNN is therefore leveraged as a dynamic model to predict each terminal state's $W^{app}$ to simplify and reduce the time consumption of DFT calculations. CGCNN has 3 convolutional layers for updating the atom feature vector $v_i$ with a nonlinear graph convolution function:

$$v_i^{t+1} = v_i^t + \sum_{j,k} \sigma[(v_i^t \oplus v_j^t \oplus u_{(i,j)_k}^t)W_f^t + b_f^t] \odot g[(v_i^t \oplus v_j^t \oplus u_{(i,j)_k}^t)W_s^t + b_s^t] \quad (3)$$

where $\oplus$ concatenate atom, neighbor, and bond features. $W$ and $b$ are convolutional weight matrix and bias of the $t^{th}$ layer which consider the interaction strength between neighbors. The $\odot$ denotes element-wise multiplication to sum neighbor atoms vector up. σ and g are the sigmoid function and softplus function, respectively, which introduce nonlinear coupling between layers. The maximum number of considered neighbors is 12. After convolution, the vectors are fed into two fully connected layers of dimensions $64 \times 128$ and $128 \times 1$ to get the final output. Details of the method can be found in the work by Xie and Grossman [45].

**1.4 Monte Carlo for TiAl/TiAlN interface generation**

The MC procedure is similar to the Miraz et al method [37] with the workflow shown in Fig. 2. The 56-atoms Ti/TiN structure is randomly doped by Al to get an initial 50 mol% of Al-doped configuration, in order to compare with RL and GA methods, the $W^{app}$ of each configuration is calculated by CGCNN or DFT no ionic update calculations, the details are shown in results and discussion section. A total of 55 steps (N = 55) is conducted, for each step, 10 trials (T = 10) are performed. In each trial, two random Ti atoms in each TiAl and TiAlN part of the initial configuration are separately exchanged with two random Al atoms to create a new configuration, followed by $W^{app}$ calculation by CGCNN or DFT no ionic update calculations. The highest $W^{app}$ among trials and initial configuration is selected as the new initial configuration for the next step. This process is repeated until the maximum number of steps is reached. The final result is the configuration with the highest $W^{app}$ obtained through the MC procedure.

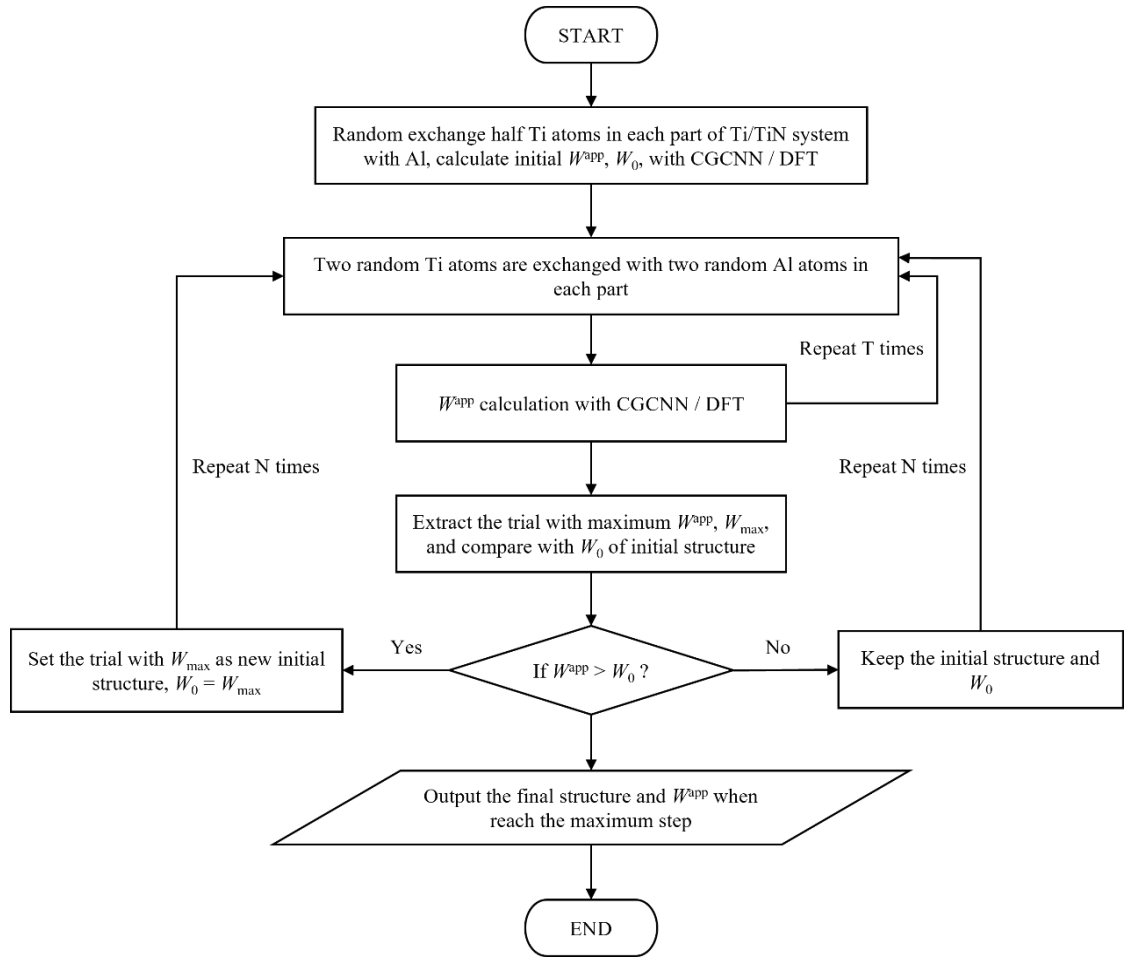

**Fig. 2.** Flow chart of the Monte Carlo (MC) procedure.

**1.5 Genetic algorithm (GA) for TiAl/TiAlN interface generation**

The workflow of the GA process is outlined in Fig. 3, and its concept is inspired by the principles of natural selection [69]. Initially, a set of 50 TiAl/TiAlN structures (N = 50) is generated randomly to form the initial population. Next, the CGCNN model is used to calculate the Wapp of these structures. Elitism selection [70] is then employed to select the first 10 structures (M = 10) with the highest $W^{app}$, which will directly pass on to the next generation. For the remaining offspring, the tournament selection method is applied to choose parents from the initial 50 structures. Specifically, 10 structures are randomly selected from the initial population, and the one with the highest $W^{app}$ is chosen as the parent. When two parents are selected, a one-point crossover is applied to generate a child. In short, there are two types of one-point crossover: the ceramic part of parent A is combined with the metal part of parent B, or the reverse, with each type having a 50% probability of occurrence. After generating the child, there is a 50% probability of mutation. A swap mutation is used to exchange one Ti atom in each part of the child with one Al atom, as illustrated in Fig. 3, where the dashed double arrows indicate the atoms being exchanged. This process is repeated 40 times (N - M = 40) and combined with the 10 parents from elitism selection to form the next generation. This new generation becomes the initial population for the next cycle, and the entire process is repeated until the number of steps reaches 9200 (S = 9,200).

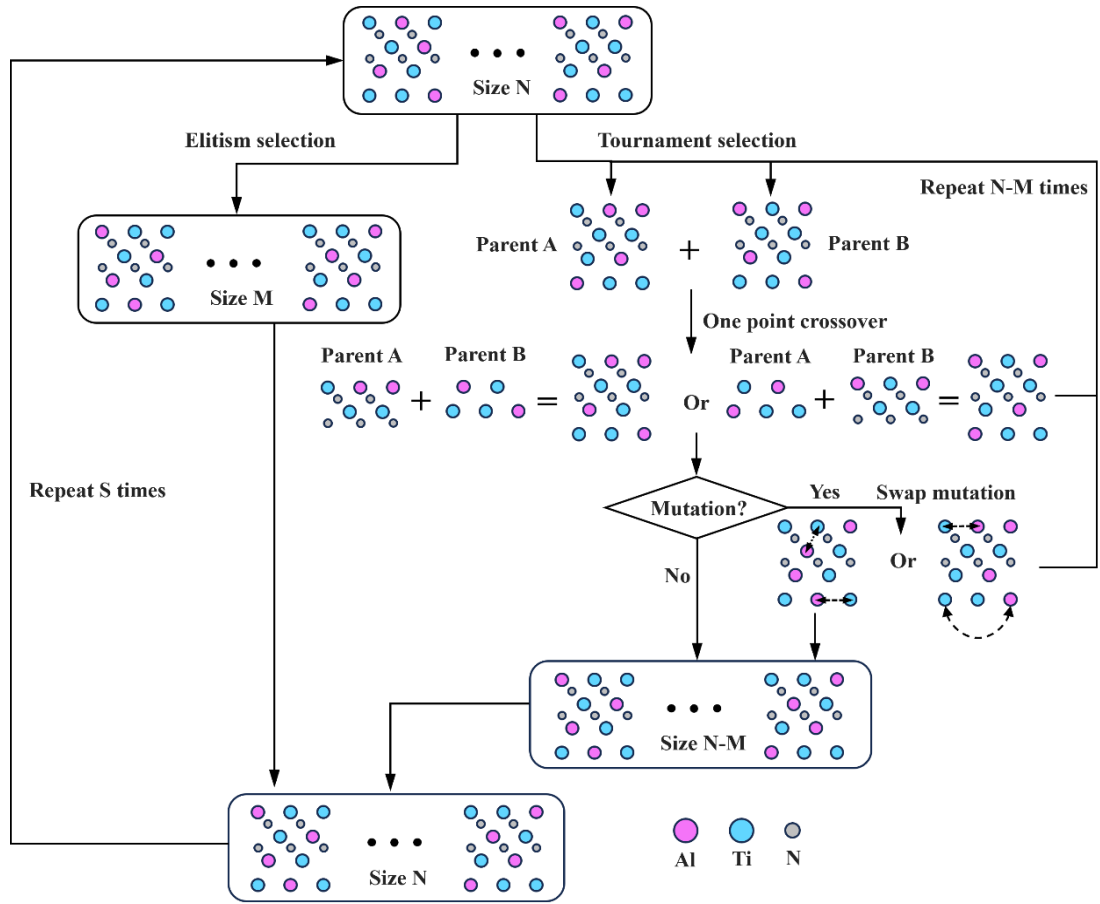

**Fig. 3.** Flowchart of the genetic algorithm (GA) procedure.

## 1.6 Reinforcement learning (DQN) for TiAl/TiAlN interface generation

Similar to the method used in the design of 2D stretchable kirigami materials [47], we use the RL model to find an optimal policy (action) that generates structures with high $W^{app}$. But we use CGCNN instead of convolutional neural network (CNN) as a dynamic model (environment) to predict each terminal state's $W^{app}$ due to consideration of period boundary in crystal. Here, each structure has 28 atomic layers, as shown in Fig. 1, so we decompose each structure into 28 atomic layers, and each layer has two atoms and four actions, each action represents one combination of the atomic layer (Ti+Ti, Ti+Al, Al+Ti, and Al+Al atoms). The action of each layer will affect the subsequent action to make sure each part composition of Al is 50 mol%. Fig. 4 shows the schematic of our model, which contains an RL agent and a dynamic model of $W^{app}$ as a policy evaluator. The RL agent formulates the construction of TiAl/TiAlN interface structures as a sequential decision-making process, where the decision regarding the combination (action) for each atomic layer is made layer by layer. The decision process from the first atomic layer to the last constitutes an episode of RL, with each episode comprising 28 actions. During an episode at each time step t, the RL agent makes an action for the $t^{th}$ layer from the 4 available actions, then in the next time step, the intermediate TiAl/TiAlN structure generated from the $t^{th}$ time step will be the input state for the RL agent. At each time step, the dynamic model calculates $W^{app}$ of each generated structure and returns the rewards to the RL agent. The reward is defined below:

$$\text{reward}(t) = \begin{cases} 0 & t < T \text{ or } t = T \text{ and } W^{app} \leq 9.5 \\ W^{app} & t = T \text{ and } W^{app} > 9.5 \end{cases} \quad (4)$$

The reward is 0 if the structure is not in a terminal state or if it is in a terminal state but the final $W^{app}$ is less than or equal to a certain threshold (set at 9.5 in this case). If the structure is in a terminal state and the final $W^{app}$ is greater than this threshold, the reward equals $W^{app}$. The choice of threshold is a critical hyperparameter that requires careful consideration. Larger thresholds can make it challenging for the RL agent to learn, while smaller thresholds may prolong the learning process. Therefore, selecting an appropriate threshold is essential for the effective training of the RL agent. DQN is a convolutional neural network (CNN), whose input is a $2 \times 28 \times 3 \times 7$ tensor that collects the intermediate state ($s_t$) of the interface structure at and before time step t in one episode and outputs the state-action value $Q(s_t, a_t)$ for all the actions available in state $s_t$. The first dimension stores information about the atomic layers of atoms being exchanged. The rest dimensions represent the TiAl/TiAlN structure, due to the atomic positions being arranged periodically, we convert the TiAl/TiAlN structure into tensor form, and each dimension represents the total number of positions in which the atoms can be, such as 28 is the number of z value (atomic layer) for an atom, 3 and 7 separately represent the number of x and y value for an atom, and different numbers represent different elements (e.g., 0 for none, 1 for N, 2 for Ti, and 3 for Al). The CNN network consists of two convolutional layers of dimensions 16 × 28 × 3 × 7, and 32 × 27 × 2 × 6 followed by three fully connected layers of dimensions 4160 × 1024, 1024 × 512, and 512 × 4. $Q(s_t,a_t)$ is an estimate of the expected reward starting from time step t to the end of the episode by a given policy. As a function approximator, DQN estimates the Q-values with parameters θ, i.e., $Q(s_t,a_t,\theta) \approx Q(s_t,a_t)$. It can learn a policy (π) that maximizes the $Q(s_t,a_t,\theta)$, Eq. 5 shows the optimal Q function as the maximum return that can be obtained starting from state $s_t$, taking action $a_t$ and following the optimal policy thereafter:

$$Q(s_t, a_t, \theta) = max_\pi E[r_t + \gamma r_{t+1} + \gamma^2 r_{t+2} + \cdots \gamma^{T-t-1} r_N] \quad (5)$$

this function obeys the Bellman optimality equation:

$$Q(s_t, a_t, \theta) = E[r_t + \gamma max_{a_{t+1}} Q(s_{t+1}, a_{t+1}, \theta)] \quad (6)$$

where the $\gamma$ is a discount factor which ensures the sum converges. In DQN, two neural networks, target net and policy net, with the same configurations are deployed to improve stability. Based on Bellman optimality equation (Eq. 6), the target net calculates the expected Q-value of the next state which is also called the temporal difference (TD) target (Eq. 7) with old parameters $\theta_{i-1}$ and compares it with the Q-value calculated by the policy net with updated parameters. The disparity between these two Q-values is commonly referred to as the Temporal Difference (TD) error, a crucial metric used in the loss function to optimize both networks. As continual learning, the DQN may abruptly lose the learned parameters from the previous task due to the weights and biases that are critical to the previous task being changed to fit the current task during learning and thus become less accurate for model. This phenomenon, called catastrophic forgetting [71, 72], is a particular challenge in artificial neural networks. To mitigate this, elastic weight consolidation (EWC) is utilized during the training of the RL agents. In short, EWC will constrain the learning of certain weights which are important for the previous task based on Fisher information matrix [73]. Here, Smooth L1 loss function combined with penalty which is summation of gradients of parameters is used to minimize the TD error, as shown in Eq. 8, where $l_t$ is Smooth L1 loss function (Eq. 9), this loss function is less sensitive to outliers and prevents exploding gradients by introducing a value $\beta$, and $\alpha$ is scale factor of penalty $P$ which is the EWC part. The details of performances by different loss functions are listed in Fig. 7a. During training, the experienced replay with reply memory of size 14500 is utilized for sample selection and expected Q-value calculations, the target net will be updated per 10 episodes. A decaying epsilon threshold (ε) for the ε-greedy strategy is used in order to relieve the exploration-exploitation dilemma [74], which means at each time step, the DQN will choose an action from the current state that maximizes Q($s_t$,$a_t$) with 1-ε probability or a random action with ε probability.

$$Q(s_t, a_t, \theta_{i-1}) = r_t + \gamma max_{a_{t+1}} Q(s_{t+1}, a_{t+1}, \theta_{i-1}) \tag{7}$$

$$\mathcal{L} = l_t + \alpha * P \tag{8}$$

$$l_t = \begin{cases} 0.5[Q(s_t, a_t, \theta_i) - Q(s_t, a_t, \theta_{i-1})]^2/\beta, & \text{if } |Q(s_t, a_t, \theta_{i-1}) - Q(s_t, a_t, \theta_i)| < \beta \\ |Q(s_t, a_t, \theta_{i-1}) - Q(s_t, a_t, \theta_i)| - 0.5 * \beta, & \text{otherwise} \end{cases}$$

(9)

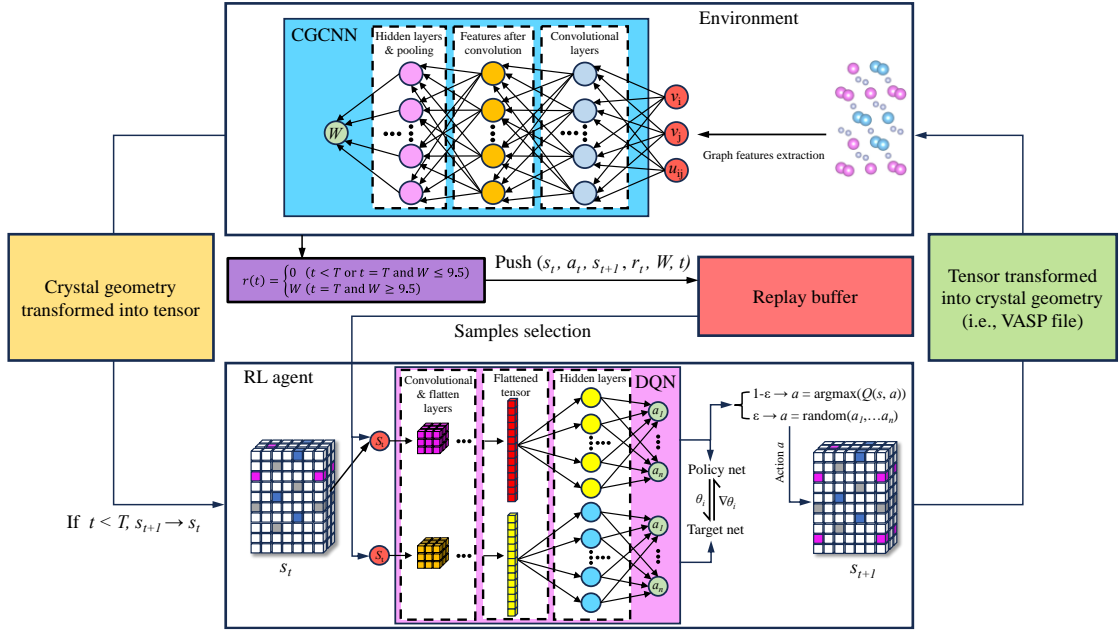

**Fig. 4.** Schematic of reinforcement learning (RL) model, which mainly consists of RL agent, environment, and replay buffer. Replay buffer stores generated structures information (state, action, next state, reward, work of adhesion, and step), RL agent selects samples from replay buffer to train and update the deep Q-network (DQN) model, and selects action to generate new structures $s_{t+1}$ based on current state $s_t$. The environment that contains crystal graph convolutional neural networks (CGCNN) potential will predict the approximate work of adhesion ($W^{app}$) and calculate reward (r) based on $W^{app}$, then push new structure information into the replay buffer for later sampling.

## 2. Results and discussion

### 2.1 Performance of CGCNN

To generate the TiAl/TiAlN structures for training the CGCNN model, the optimal initial Ti/TiN system needs to be defined for Al atoms exchange. Three initial types of interfacial stacking sequences, FCC-like stacking, HCP-like stacking, and node-to-node stacking, with 7 different distances from 0.5 Å to 3.5 Å between interfaces in each type are considered. Fig. 5a illustrates the three types of stacking, the top row shows the 3D view of the three layers of interfaces consisting of 2 bottom layers of TiN and one top layer of Ti. The second row in Fig. 5a shows the details of these three types of stacking sequences, the stacking sequences at the interfaces (dashed line) of FCC-like, HCP-like, and node-to-node are ABC, ABA, and ABB, respectively. For each type of stacking, seven different interlayer distances between the metal and ceramic surfaces are set, and DFT is used to calculate the total free energy of each structure to determine the most stable initial Ti/TiN structure. The total free energy of each structure is shown in Fig. 5b. It is evident that the node-to-node stacking (red square) has the highest total free energy, indicating greater instability compared to the others. The inset in Fig. 5b shows the total free energy of both FCC-like (orange circle) and HCP-like (cyan triangle) stackings under 1 Å and 1.5 Å interlayer spacing. Since FCC-like stacking with 1 Å interlayer spacing has the lowest total free energy and small mismatch, it is chosen as the basic structure for Al doping (see Fig. 1).

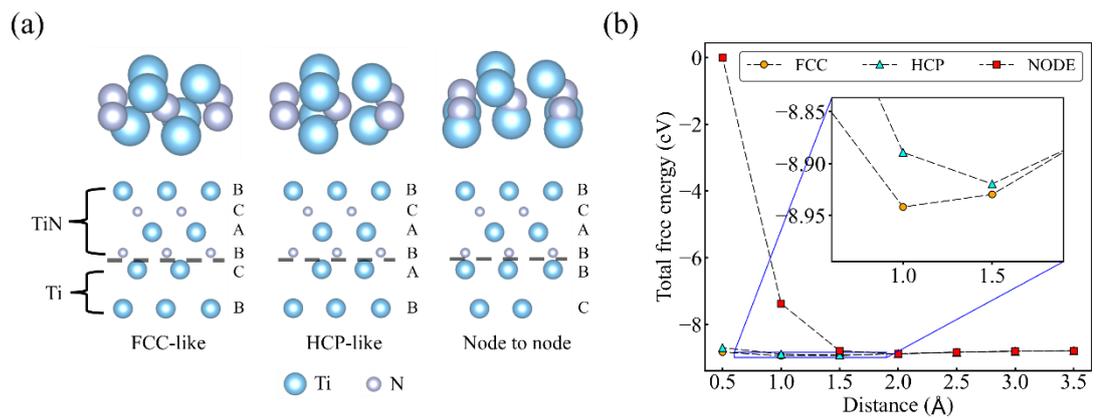

**Fig. 5.** (a) Three types of stacking at the interface: FCC-like, HCP-like, and node-to-node. (b) Total free energy as a function of interlayer distances between metal and ceramic surfaces, it can be seen from the inset that FCC-like stacking with 1 Å interlayer spacing has the lowest total free energy.

A total of 2000 TiAl/TiAlN structures, each with 50 mol% Al dopants in both TiAl and TiAlN systems, are randomly generated based on basic Ti/TiN structure and get their $W^{app}$ (Eq. 2) by DFT static energy calculations as our training dataset. Fig. 6a is the histogram representing the distribution of the $W^{app}$ in these 2000 structures, this distribution is highly sparse, where only 0.1% (2 out of 2000) structures have $W^{app}$ higher than 11.4 eV. Therefore, it is difficult to find structures with high $W_{ad}$ by using random search. To choose the best model, a train-validation scheme is utilized to optimize the prediction of $W^{app}$ of TiAl/TiAlN structures. Each model is trained with part of the 2000 training data and then validated with part of the rest data, and the best-performing model in the validation set is selected. Here, the mean absolute error (MAE) of $W^{app}$ per atom (eV/atom) under different numbers of training data is shown in Fig. 6b. 5 different sizes of training set, 160, 300, 520, 1000, and 1500, are separately used to train the CGCNN model, it is obvious that the MAE approaches to constant after 500. Furthermore, Fig. 6c shows the $W^{app}$ of the 2,000 structures calculated by DFT versus the predicted values by CGCNN. Most of the points are located on the black dashed line, which represents the ideal fit (1:1). Therefore, the CGCNN model, trained with 520 data points, is utilized as our dynamic model to train the DQN model.

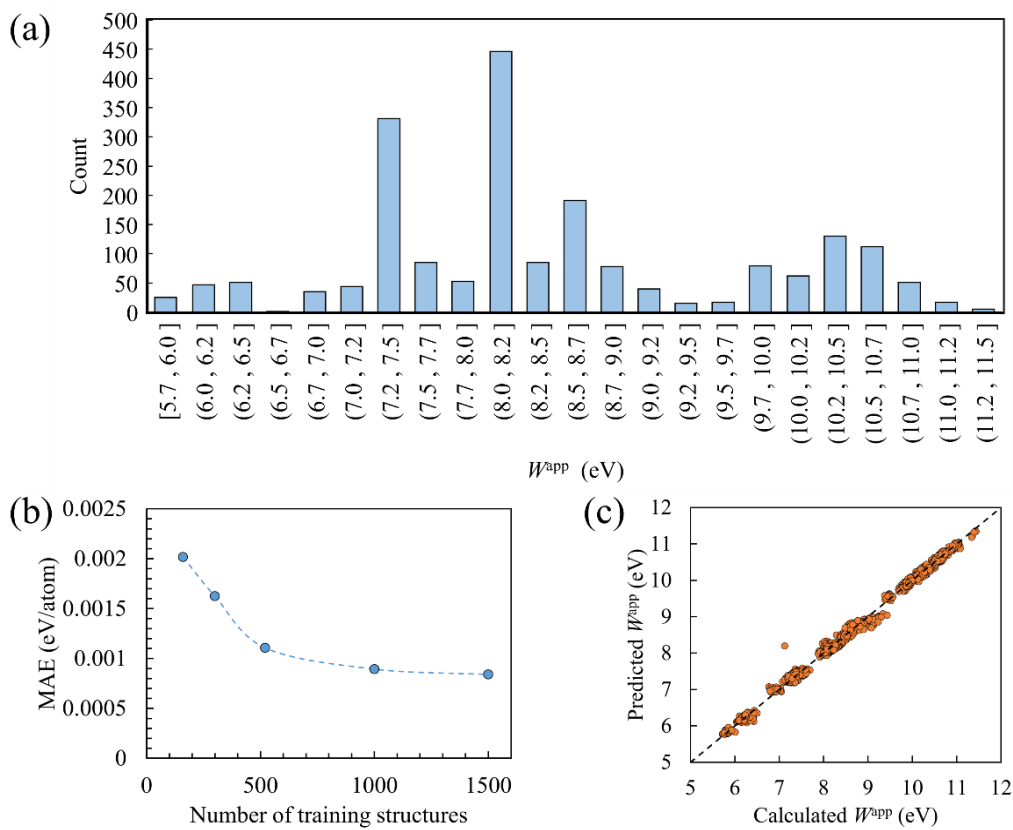

**Fig. 6.** Performance of crystal graph convolutional neural networks (CGCNN) model. (a) Histogram representing the distribution of the approximate work of adhesion ($W^{app}$) in the training dataset. (b) Mean absolute error (MAE) as a function of training data size for predicting $W^{app}$. (c) The $W^{app}$ of the 2000 structures calculated by DFT versus predicted values by CGCNN, the black dashed line represents the ideal fit.

**2.2 Performance of DQN**

By interacting with the CGCNN model, the DQN can update each state-action pair's Q-value based on the $W^{app}$ (reward) of the current intermediate structure (state) predicted by the CGCNN model during training. Finally, when there is an intermediate structure, the trained DQN can select the combination (action) that can receive the highest Q-value as the next optimal action, by repeating this process until the final structures with high $W^{app}$ are obtained. Here, a total of 2000 training episodes are applied in DQN, and the mean $W^{app}$ of DQN and random search are evaluated during training. The mean $W^{app}$ is the average $W^{app}$ of 50 structures generated by a target net of DQN per 10 episodes. We also use a random search method to generate 50 structures per 30 episodes and calculate their $W^{app}$ by trained CGCNN model. The mean $W^{app}$ is shown in Fig. 7a, to compare the performance of DQN with different loss functions, three DQN models are illustrated here, all with a set threshold of 9.5 eV, updating the target net value every 10 episodes, and a batch size of 50, except for using a different loss function. It can be seen with the rise of training episodes, the DQN will learn the output of the CGCNN model (environment) and optimize the final structure to get the higher $W^{app}$, thus the mean $W^{app}$ increases during the training. The random search, on the other hand, has no ability to capture the information of the CGCNN model, thus the mean $W^{app}$ is oscillated around the highest probability of occurrence (Fig. 6a). Compared among the DQN models, the DQN model with the EWC has highest outputs, since EWC can slow down the learning process and keep the parameters' importance of previous tasks, which leads to more stable and higher outputs. On the other hand, due to catastrophic forgetting, the performance of the DQN model without EWC decreases when trained between 1000 and 1500 episodes, which makes the model less accurate in predicting training samples. After training, the DQN follows an ε-greedy policy to generate structures. The performance of trained DQN under different ε values is shown in Fig. 7b. 10 different ε are used to create 2000 structures in each ε, the total time consumption, maximal $W^{app}$, and the number of structures with $W^{app}$ higher than 11 eV are extracted. With the rise of ε, the probability of selection of random action increased, and both max $W^{app}$ (red square) and the number of high $W^{app}$ structures (cyan circle) decreased, this trend is consistent with Fig. 7a and has been explained previously.

When the ε increases, the time consumed (orange triangle) is reduced linearly, this is due to the random search not calling the policy net to select action, which requires extra time for outputting Q-values. While the time consumption is increased when using lower ε (from 89s to 133s), the number of created potential structures is drastically increased (from 29 to 432) compared to the time consumption, so we use trained DQN model with 0.1 ε to generate structures in the following work. Fig. 7c displays the performance MC method for generating high $W^{app}$ structure with DFT no ionic update calculations, here to consist with the training size (520) of the RL model, a total of 55 steps (550 structures) are generated by the MC method. To connect the MC method with DFT calculation, Atomic Simulation Environment (ASE) and VASP are used [75]. Two tests with different initial configurations are conducted to check if the MC method can yield similar final configurations. It can be seen that the two tests have similar results, both smaller than that predicted by DQN (black dashed line), indicating that under the same data size, DQN can find optimal structure more efficiently.

Additionally, instead of using DFT calculations to find optimal structures, like RL and GA methods, the MC method can also be used in the CGCNN model to get optimal structures, as shown in Fig. 2. A comparison of different methods is shown in Table 1. In Table 1, under the same time consumption, the DQN with combined loss function and GA methods have both the highest $W^{app}$ prediction, which is 15.552 eV, but DQN has a higher prediction in total. On the other hand, the MC method had the lowest predictive value in both tests. What is more noteworthy is that for the number of high $W^{app}$ structures, the MC method has few or even only one structure among the 384,000 generated structures, while DQN and GA have more high-$W^{app}$ structures in 70,000 and 368,050 generated structures respectively, and both have high deviation for count of high $W^{app}$ structures. Even though, DQN and GA methods can generate more candidate structures than MC methods under the same time consumption.

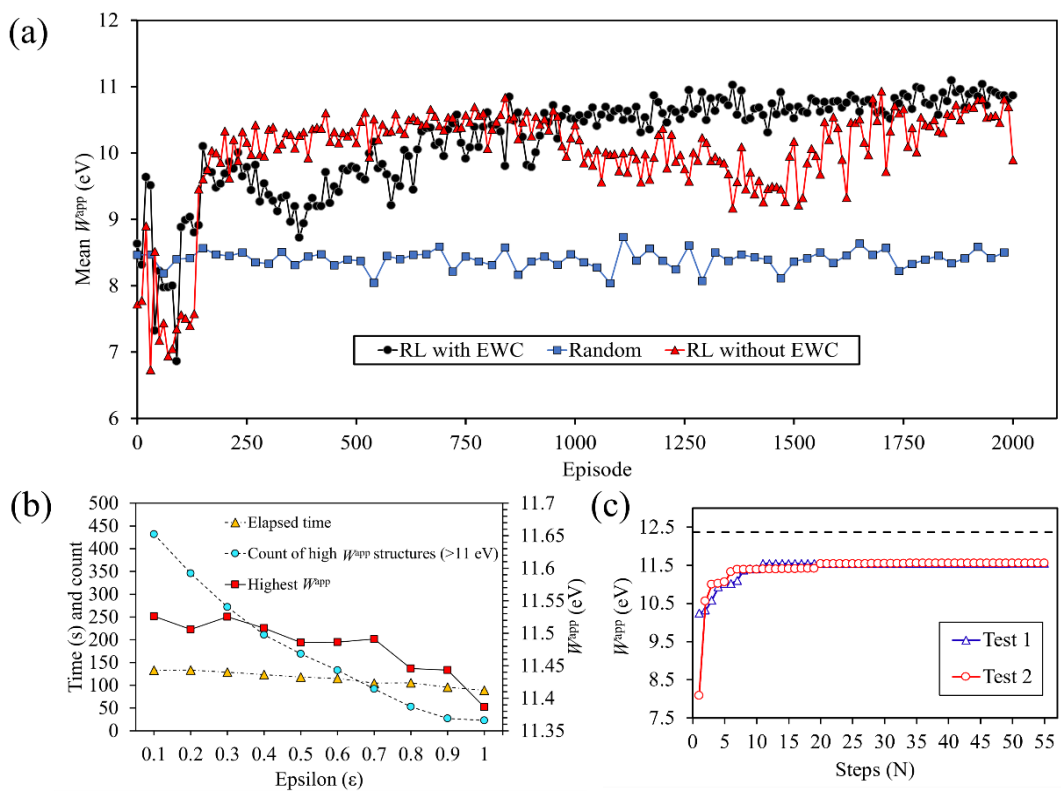

**Fig. 7.** Performance of deep Q-network (DQN) model. (a) The evolution of mean approximate work of adhesion ($W^{app}$) of generated structures by DQN models with different loss functions and random search during training. (b) The time consuming of trained DQN under different ε values. (c) $W^{app}$ versus steps in the Monte Carlo (MC) method.

**Table 1.** Performance of deep Q-network (DQN), genetic algorithm (GA), and Monte Carlo (MC) methods.

| Methods | DQN test 1 | DQN test 2 | GA test 1 | GA test 2 | MC test 1 | MC test 2 |
| --- | --- | --- | --- | --- | --- | --- |
| Highest $W^{app}$ (eV) | 11.552 | 11.535 | 11.552 | 11.520 | 11.511 | 11.504 |
| Count of high $W^{app}$ (>11.5 eV) | 95 | 20 | 337 | 28 | 4 | 1 |

**2.3 TiAl/TiAlN interface generation and work of adhesion**

We apply the CGCNN model-based RL agent and random search to generate 100000

and 350000 TiAl/TiAlN structures which have different $W^{app}$. In these structures, 60 and 12 different TiAl/TiAlN structures with $W^{app}$ higher than 11.5 eV are generated in RL agent and random search separately. The $W^{app}$ calculated by both DFT and CGCNN models of these 72 structures and extra 800 structures in different $W^{app}$ ranges are plotted in Fig. 8a. In general, the predictions fluctuate above and below the DFT values, which indicates the good performance of the CGCNN model. For high $W^{app}$, however, the predictions are lower than that of DFT, as shown in the inset of Fig. 8a. As the $W^{app}$ increases, the predicted value of CGCNN gradually approaches an upper limit, causing its predicted value to oscillate around 11.5 (green dotted line). This systematic error is determined by the limitation of the training set. The upper limit of $W^{app}$ in training data is less than 11.5 (11.43 eV), which results in the model's extrapolation capability being limited to around 11.5. Even so, the existing systematic errors are small and there are no cases of excessive errors, and the overall growth trends of DFT and predicted values are similar, this indicates the graph-based features in CGCNN capture some local environment information of each atom, this enables the CGCNN model to have certain extrapolation capabilities under similar structures. Even if the CGCNN model underestimates the actual $W^{app}$, its predicted value still falls within the highest energy range in the DFT calculation. Therefore, we select structures with a predicted energy greater than 11.5 as our analysis objects. Additionally, since the CGCNN is trained on structures without relaxation to improve efficiency, we use Eq. 2 to calculate the $W^{app}$ of unrelaxed structures. This can lead to discrepancies with the relaxed value $W_{ad}$ (Eq. 1). Therefore, to examine if the $W_{ad}$ has a similar energy distribution with $W^{app}$, the $W_{ad}$ and $W^{app}$ of the 20 structures with high $W^{app}$ (18 for RL agent and 2 for random search) and extra 27 random structures are calculated by DFT as well. Fig. 8b plots the $W_{ad}$ and $W^{app}$ calculated by DFT values, we can observe that although the overall $W_{ad}$ is smaller than $W^{app}$, they have a similar overall change trend. The cyan and green dashed lines are the $W^{app}$ and $W_{ad}$ of unrelaxed Ti/TiN and relaxed Ti/TiN correspondingly, some parts of $W^{app}$ and $W_{ad}$ of TiAl/TIAlN are higher than that of Ti/TiN, which shows the TiAl/TiAlN has stronger bonds at the interface. The structures within the black dotted box are the high $W^{app}$ structures predicted by both RL agent and random search and their corresponding $W_{ad}$ after relaxation, we can see the structures that have $W^{app}$ higher than 11.5 eV also have higher $W_{ad}$ than others, this also confirms that it is feasible to use the $W^{app}$ of structures without relaxation to train our model and predict high $W_{ad}$ structures.

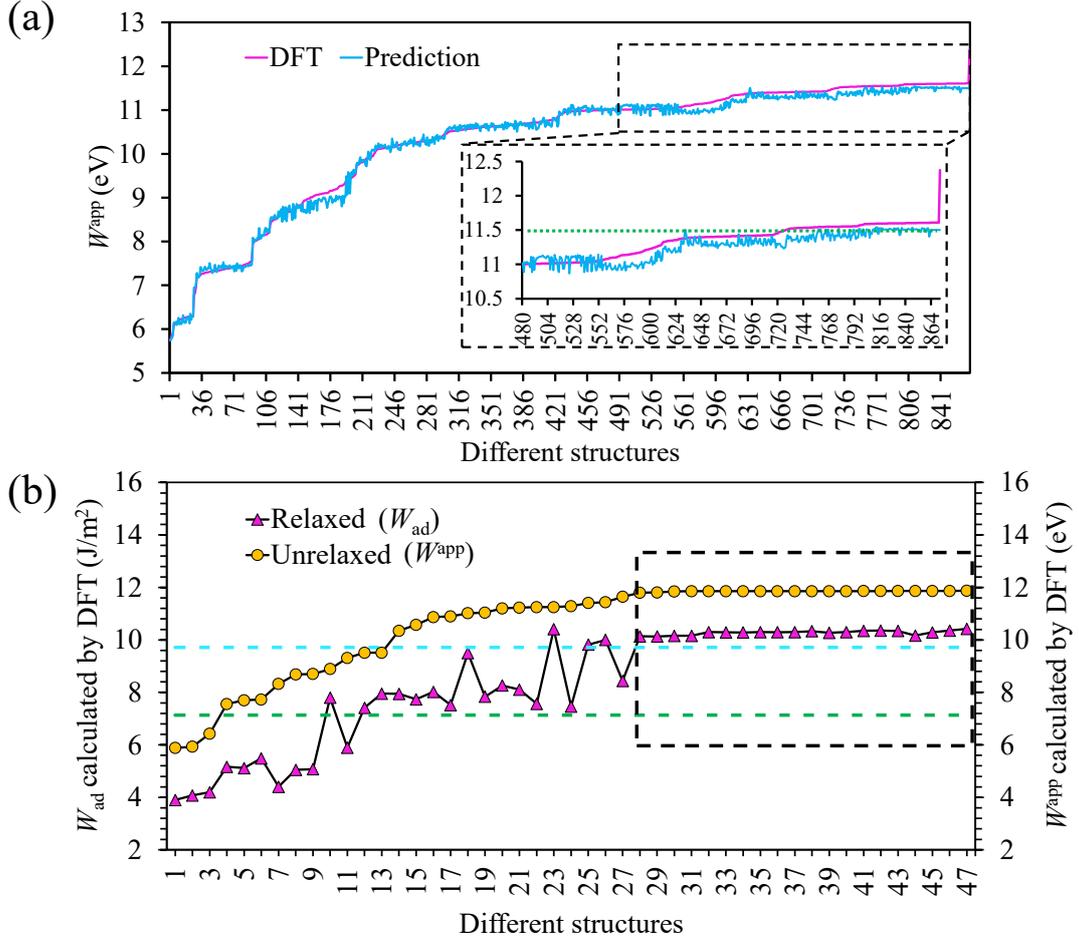

**Fig. 8.** (a) The approximate work of adhesion ($W^{app}$) calculated by both DFT and crystal graph convolutional neural networks (CGCNN) model of 72 predicted high $W^{app}$ structures (60 for reinforcement learning (RL) agent and 12 for random search) and extra 800 structures in different $W^{app}$. (b) The work of adhesion ($W_{ad}$) and $W^{app}$ of the 20 structures with high $W^{app}$ (18 for RL agent and 2 for random search) and extra 27 random structures.

## 2.4 Interpretability of CGCNN and RL models

To further investigate the structures with high $W^{app}$, both CGCNN and RL models are interpreted using principal component analysis (PCA) to gain insights. In the CGCNN model, two vectors, local environment representation and averaged total representation, are analyzed. The local environment representation is the output of the convolutional layers, when the vector of input which contains the atoms and bonds information passes through $t$ convolutional layers, the convolution operations (Eq. 3) will make the vector have the local environment information such as atomic types and geometries. Therefore, the local environment representation can describe the similarity between the local environments of any two structures. The averaged total representation is the output vector obtained by adding a pooling layer (Eq. 10) after the convolutional layer.

$$v = (\sum_i v_i^t)/n \tag{10}$$

This process is analogous to the total energy in classical potentials, which is calculated by summing the local atomic energies of all atoms in the system. Here, the averaged total representation contains the contribution of each local chemical environment and conveys information about the structure. For the RL model, since the policy network can output the Q-value based on the current state and next action, the Q-value of each state with different subsequent actions can be calculated by the trained RL model and visualized using PCA.

Fig. 9a and 9b show the PCA of local environment representation and averaged total representation of the CGCNN model, respectively. A total of 2072 structures consisted of train data and the 72 high $W^{app}$ structures predicted by RL and random search methods are plotted. For local environment representation, since each structure has 56 atoms, the local environment representation for each structure is a 56 × 64 matrix. Therefore, before applying PCA, the local environment representation is flattened from a 56 × 64 matrix to a 1 × 3,584 vector. The averaged total representation reduces the 56 × 64 matrix to a 1 × 64 vector by element-wise addition of each local environment representation (Eq. 10). After PCA, the original 3584 and 64-dimensional vectors are reduced to 2 dimensions. In Fig. 9a and 9b, the structures with similar $W^{app}$ are clustered together. Except for the cluster with the highest $W^{app}$ (since these structures come from the 72 predicted structures that are out of training data), the distribution of the rest clusters is similar to the histogram of training data (Fig. 6a), and the energy is gradually changed in one direction, all indicate that the CGCNN model embeds the physical meaning such as the geometry of the structure, atomic information, and local environment of the atom into each representation. By further inspecting the structures of both high and low $W^{app}$, some specific patterns for these two types of structure, especially around the interface of TiAl/TiAlN are found. As shown in Fig. 9b, two examples of interface for high $W^{app}$ structure and low $W^{app}$ structure are illustrated in red dashed inset and blue dashed inset, respectively. Total 10 atomic layers, where top 6 layers come from TiAlN and bottom 4 layers come from TiAl, are extracted in each structure, it can clearly see that the metal layers ($5^{th}$ and $7^{th}$ layer from top to bottom) on both sides of the N atomic layer ($6^{th}$ layer from top to bottom) at the interface exhibit an inverse pattern in these two types of structures, where the element of the metal layer is Al and the ceramic layer is Ti for high $W^{app}$ structures, on the contrary, the element of the metal layer is Ti or Ti + Al and the ceramic layer is Al for low $W^{app}$ structures, for the other structures, their interface have the patterns other than these patterns. Furthermore, the AlN structure mainly appears at the interface of high $W^{app}$ structures, and an Al layer follows the Ti layer at the interface of the TiAl part. Fig. 9c shows the 2D Q-values when the ceramic layer generation is completed, and the metal contact layer is about to be generated (as shown in the inset of Fig. 9c). Here, a 1 × 512 Q-value vector predicted by the middle-hidden layer for each structure are utilized, and each vector is reduced to 2D by PCA. The Q-values larger than 3 and smaller than 1.5 are plotted, a clear separation can be seen in these Q-values, and by comparing to the structures and actions in both high and low Q-values, we can notice that for high Q-value structures, their following actions mainly generate the Ti atomic layer, and their previous non-N atomic layer is the Al layer, for low Q-value structures, they are just the opposite of the high Q-value structures, which is consistent with the two patterns found previously. Besides the interpretability of CGCNN and RL models, the similarity of each structure compared to others is also studied. To simply calculate the similarity between any two structures, the tensor form (28 × 3 × 7) of each structure is used. To determine the correlation between two structures, the dot product of the corresponding row vectors (1 × 7) from the structure tensor of each structure is calculated. This process is similar to calculating the cosine similarity between two matrices. However, the cosine similarity between any two structures is always 1 because the corresponding row vectors are in the same direction. Therefore, the magnitude of each vector is ignored to account for the influence of atom type. Then, element-wise summation is applied to each tensor to reduce the dimension from 28 × 1 × 3 to 1 × 3. We repeat these processes until each structure is compared to all structures, including itself. After the flattening process, the final similarity vector of size 1 × 6216 for each structure is obtained. This vector indicates how similar a structure is compared to other structures. Fig. 9d is the 2D similarity of each structure, the high $W^{app}$ structures are clustered at the top region, though other structures are more sparsely scattered across the figure due to the larger structure space. This indicates that there are more similarities between high $W^{app}$ structures' vectors, in other words, the high $W^{app}$ structures are similar, which also explains why the values of high $W^{app}$ or $W_{ad}$ have similar values (Fig. 8b black dotted box). Based on the above analysis, we conclude that the RL model learns the structures and develops a method that can generate high $W^{app}$ structures. These structures follow a certain pattern that leads them to have high.

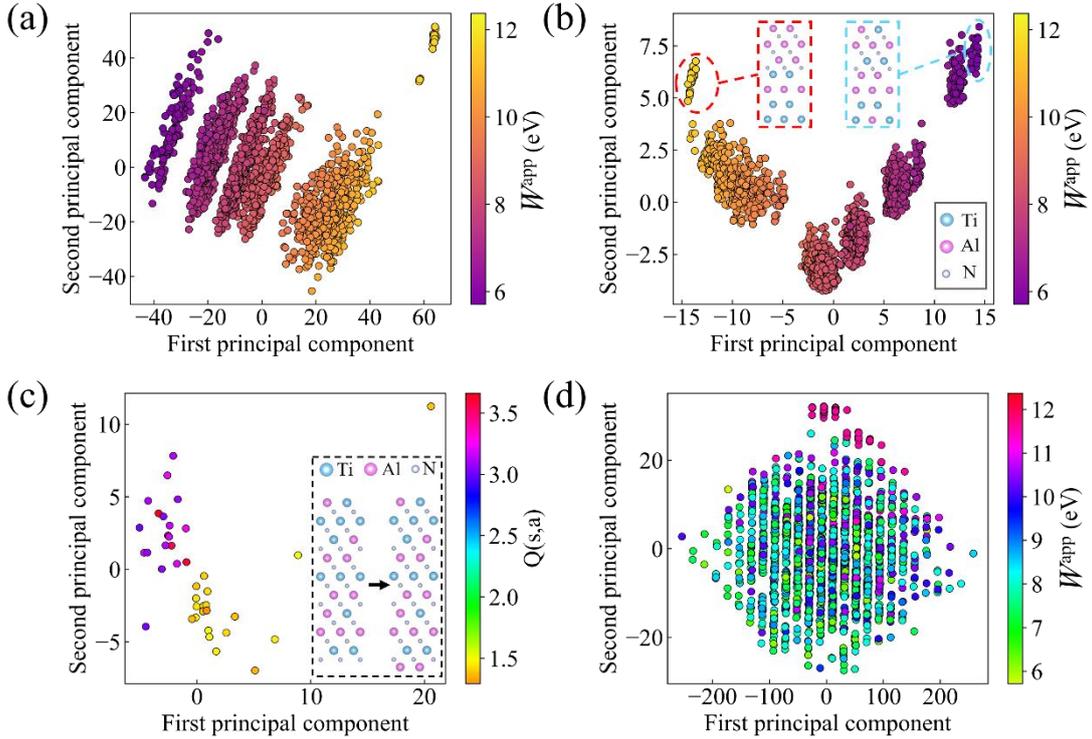

**Fig. 9.** Visualization of different feature vectors by principal component analysis (PCA). (a) PCA of output vectors of the convolutional layers in crystal graph convolutional neural networks (CGCNN). (b) PCA of output vectors of the pooling layer in CGCNN. (c) PCA of output vectors of middle hidden layer in deep Q-network. (d) The similarity vector of each structure compared to others is visualized by PCA.

**2.5 Electronic structure and bonding**

To understand the strengthening mechanism of the interface, we calculate the charge density of the TiAl/TiAlN and Ti/TiN by DFT. And the information on bonding and electronic structure of the interface can be obtained by charge density difference. The charge density difference is calculated by:

$$\Delta\rho = \rho_{all} - \rho_{metal} - \rho_{ceramic} \tag{11}$$

where $\rho_{all}$ is the total charge density of the TiAl/TiAlN or Ti/TiN system, $\rho_{metal}$ and $\rho_{ceramic}$ are the charge density of the corresponding metal and ceramic parts, respectively.

The charge density difference at the interface of the TiAl/TiAlN and benchmark Ti/TiN structures are investigated here. Fig. 10a, b, and c show the 2D charge density difference on $(\bar{1}\,\bar{1}\,2\,0)/(1\,\bar{1}\,0)$ plane of the TiAl/TiAlN interfaces with high $W_{ad}$ (10.34 J/m$^2$), low $W_{ad}$ (3.90 J/m$^2$), and Ti/TiN (7.14 J/m$^2$) interface, respectively. The top purple dashed line represents the surface of the ceramic, and the bottom represents the surface of the metal. The red and blue areas represent electron accumulation and depletion, respectively. In these figures, some atoms are numbered for identification. From these figures, we can see that atom 4 at the metal-ceramic interface transfers charges to atom 3 at the ceramic interface and forms a bond. For high $W_{ad}$ and Ti/TiN structures, this bond exhibits ionic characteristics, while for low-energy structures, this bond is more like a polar covalent bond. The different features of the bond between these two types of bonds can be explained by the bond strength since the Ti-N bond (476.1±33.1) is stronger than the Al-N bond (297±96) [76]. In the case of high $W_{ad}$ structure (Fig. 10a), aluminum atom 2 transfers charge to atom 1, atom 4, and atom 8, thus forming bonds or interactions with these atoms. Furthermore, we can observe that atom 1 that gains charge also interacts with atom 4 at the metal interface. The charge distribution in the ceramic part of Ti/TiN is similar to the high $W_{ad}$ structure, and atom 4 also interacts with atom 1. It's worth noting here that since the size of the aluminum atom is smaller than that of the titanium atom, when the atom 2 is aluminum, i.e., for high $W_{ad}$ structure, the distance between the upper and lower layers will be smaller compared to when the atom 2 is Ti. In Fig. 10a, b, and c, the distances between atom 1 and atom 4 are 3.61 Å, 3.77 Å, and 3.64 Å, respectively, so the interaction between these two atoms in the high $W_{ad}$ structure is easier to form. For the low $W_{ad}$ structure, titanium atom 2 increases the distance between atoms 1 and 4, and the accumulation of charge on atom 2 and the charge deficiency of atom 4 make the interaction among atoms 1, 2, 3, and 4 weak and difficult. In the metal part, atom 4 also interacts with the neighboring metal layers, which can be seen from the figures. For high $W_{ad}$ and Ti/TiN structures, atom 4 interacts with atoms 5 and 7, and the interaction is stronger in the high $W_{ad}$ structure. In addition, atom 6 in the high $W_{ad}$ structure also has a long-range effect on atom 4. Similarly, because atom 5 and atom 7 are aluminum, the distance between atom 4 and atom 6 in the high $W_{ad}$ structure is the shortest (4.55 Å), while the distances for the low $W_{ad}$ structure and Ti/TiN structure are 4.85 Å and 4.66 Å, respectively. For the high $W_{ad}$ structure, the half of 4.55 Å Ti-Ti distance is closest to the van der Waals radius of titanium element [77], so atom 6 also has a weak interaction with atom 4. Furthermore, for the high $W_{ad}$ structure, atom 5 and atom 7 have long-range effects on atom 3 and atom 8, respectively. For the low-energy structure, due to the displacement of the metal layers, atom 4 mainly interacts with atom 7, resulting in weak interactions between the metal interface and its internal layers. Fig. 10d to f illustrates the plane-averaged charge density difference along the z-direction ([0 0 0 1]/[1 1 1] direction) calculated by VASPKIT [78]. As in Fig. 10a to c, the region between the purple dashed lines is the interface. It can be seen that charges transfer from the metal part (bottom dashed line) to the ceramic part (top dashed line), and the main charge rearrangement occurs at the interface, where the accumulation and depletion peaks are also located. The peaks of accumulation (depletion) for high $W_{ad}$ structure, low $W_{ad}$ structure, and Ti/TiN structure are 4.27 (-3.58) e/Å, 1.94 (-2.98) e/Å, and 2.97 (-2.88) e/Å, respectively. The skewed and small peaks in the low $W_{ad}$ structure can also confirm that the interactions among atoms 1, 2, 3, and 4 are weak. The charge accumulation on atom 2 and the charge deficiency of atom 4 leads to less accumulation around the surface of the ceramic. Thus, the high $W_{ad}$ structure exhibits the largest charge accumulation and depletion peaks, indicating a strengthening of the interfacial interaction and charge transfer compared to the other two structures.

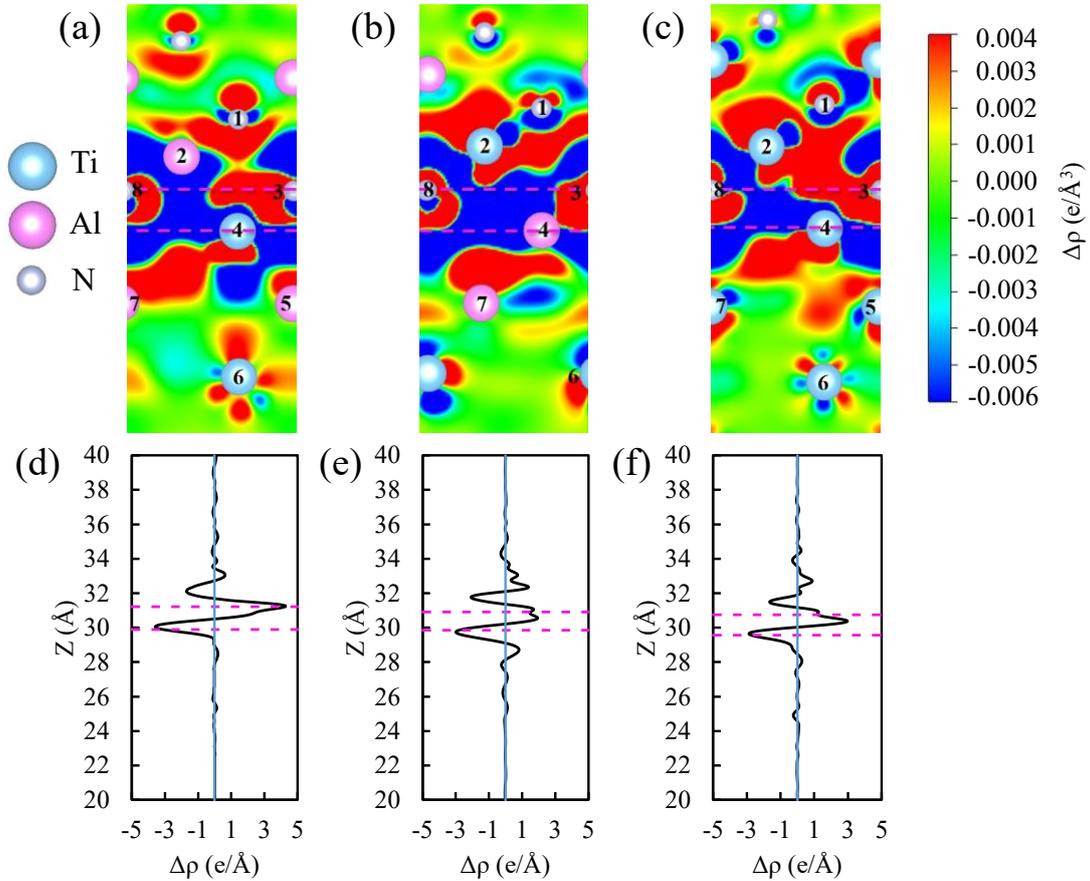

**Fig. 10.** (a) Charge density difference of TiAl/TiAlN interface with high work of adhesion ($W_{ad}$) on $(\bar{1}\bar{1}2\,0)/(1\,\bar{1}\,0)$ plane. (b) Charge density difference of TiAl/TiAlN interface with low $W_{ad}$ on $(\bar{1}\bar{1}20)/(1\bar{1}0)$ plane. (c) Charge density difference of Ti/TiN interface on $(\bar{1}\bar{1}20)/(1\bar{1}0)$ plane. (d) Plane-averaged charge density difference of TiAl/TiAlN interface with high $W_{ad}$ along $[0\,0\,0\,1]/[1\,1\,1]$ direction. (e) Plane-averaged charge density difference of TiAl/TiAlN interface with low $W_{ad}$ along $[0\,0\,0\,1]/[1\,1\,1]$ direction. (f) Plane-averaged charge density difference of Ti/TiN interface along $[0\,0\,0\,1]/[1\,1\,1]$ direction.

To quantitatively compare the interface charge transfer, the Bader charge is used [79-82]. The average charge as a function of the atomic layer at the interface of different structures is plotted in Fig. 11. The left and right purple dash-dotted lines represent the surfaces of the ceramic and metal, respectively. In the ceramic, the electronegative N atoms accumulate negative charge, while the Ti or Al atoms have a positive charge. In the metal, layer 19 has a positive charge in all structures. In layer 20, both high $W_{ad}$ and low $W_{ad}$ structures have a negative charge, although the Ti/TiN structure has a slightly positive charge. The result of Ti/TiN is the same as the found of Miraz et al [37], they think that at least two layers in the Ti-phase will be influenced by the electronegative N layer, which can weak the interactions and generalized stacking fault energy (GSFE) barrier between these two layers. This phenomenon can also be observed in Fig. 10c where atoms 3 and 8 have an effect on atoms 4, 5, and 7. They also noticed that incorporating slightly more electronegative Al atoms in layer 20 can decrease the average charge of this layer since Al and Ti atoms will be more attractive to each other than Ti and Ti atoms [83]. Thus, it can increase the GSFE barrier. Interestingly, this doping method is also applied by our RL model, which uses the same approach to generate the high Wad structure. As shown in Fig. 10a, atoms 5 and 7 are Al atoms.

This method can lead to a more negative average charge in layer 20, thus it can have more attraction to layer 19, which leads to layer 19 having a more positive average charge. It can be seen that in layers 18 and 19, both high $W_{ad}$ and low $W_{ad}$ structures have more negative or positive average charges than that of Ti/TiN, especially for low $W_{ad}$, which has the largest positive average charge in layer 19. Even though, due to the lower average charge in layers 17, 20, and 21 compared to that of the high $W_{ad}$ structure, especially for layer 17, the average charge is similar to that of the Ti/TiN structure. Therefore, the bonding strength between layer 17 and layer 18, layer 19, and layer 20 are lower than that of the high $W_{ad}$ structure. Compared to the high $W_{ad}$ structure and the Ti/TiN structure, when the atoms in layer 19 of both structures are Ti, the high $W_{ad}$ structure generally has a more negative or positive average charge around the interface. Thus, the high $W_{ad}$ structure has higher bonding strength.

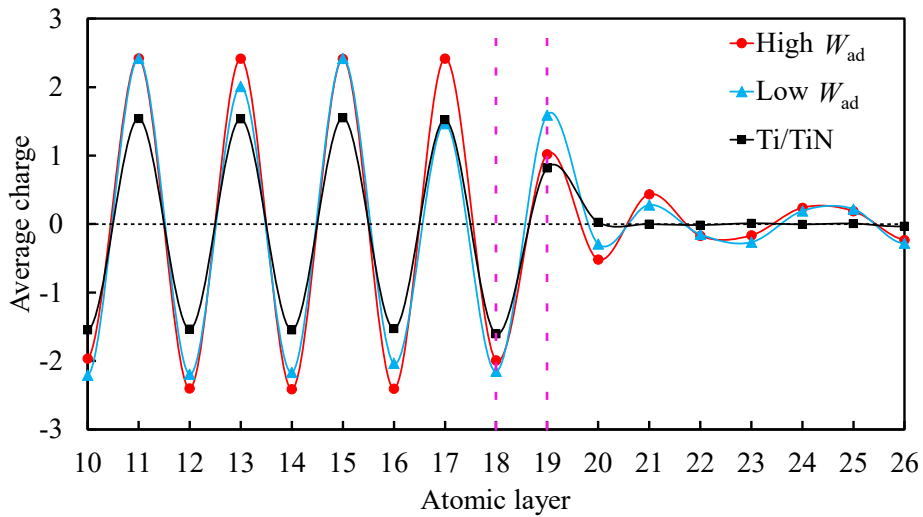

**Fig. 11.** The average charge as a function of the atomic layer at the interface of different structures.

### 3. Conclusions

We have developed a model that combines DQN with CGCNN potential to identify aluminum-doped TiAl/TiAlN interfaces containing 50 mol% aluminum. These interfaces exhibit a higher $W_{ad}$ compared to the Ti/TiN counterpart. When compared to the random search method, the RL model is more efficient and can identify TiAl/TiAlN interfaces with a higher $W_{ad}$ than those found using random search or the MC method. Furthermore, visualizations of features from the DQN and CGCNN reveal that structures with high $W_{ad}$ exhibit a specific pattern of Al atom doping near the interface. This pattern includes alternating layers of Al and N on the ceramic surface, along with a single Ti layer followed by a single Al layer on the metal surface. Further analysis of the electronic structure and bonding in high $W_{ad}$, low $W_{ad}$, and Ti/TiN structures reveals that, for the TiAl/TiAlN interface with high $W_{ad}$, the smaller size of Al atoms reduces the distance between the Ti layer on the metal surface and both the second nearest neighbor N layer and the second nearest neighbor Ti layer in the metal. This reduction in distance enhances interactions, such as van der Waals forces, between these layers, resulting in stronger bonding at the interface. Bader charge analysis shows that the presence of Al atoms in the second layer of the metal part can lead to more intense interactions with the surface Ti layer, thereby increasing bond strength at the interface. While further research is needed to refine the methods used in this study, employing machine learning potentials to approximate or replace DFT calculations offers significant promise as a fast preliminary screening method for other optimization problems with large configuration spaces. By integrating RL, structural correlations and features can be identified from these potential configurations.

**Declaration of Competing Interest**

The authors declare that they have no known competing financial interests or personal relationships that could have appeared to influence the work reported in this paper.

**Acknowledgements**

Funding for this research was provided by National Science Foundation (NSF) under award numbers, DMR-2239216, CMMI-1826439, and CMMI-1825739. The authors also thank the Agave Computer Cluster of ASU for providing the computational resources.